\documentclass[conference]{IEEEtran}
\IEEEoverridecommandlockouts
\pdfoutput=1
\usepackage{cite}
\usepackage{amsmath,amssymb,amsfonts}
\usepackage{algorithmic}
\usepackage{graphicx}
\usepackage{textcomp}
\usepackage{xcolor}
\def\BibTeX{{\rm B\kern-.05em{\sc i\kern-.025em b}\kern-.08em
    T\kern-.1667em\lower.7ex\hbox{E}\kern-.125emX}}

\usepackage{svg}

\usepackage[caption=false,font=footnotesize,labelfont=rm,textfont=rm]{subfig}
\usepackage{textcomp}

\usepackage{cite}
\usepackage{extarrows}
\usepackage{graphicx}

\def\BibTeX{{\rm B\kern-.05em{\sc i\kern-.025em b}\kern-.08em
    T\kern-.1667em\lower.7ex\hbox{E}\kern-.125emX}}

\usepackage{extarrows}
\usepackage[colorlinks,urlcolor=blue,linkcolor=blue,citecolor=blue]{hyperref}

\usepackage{utfsym}

\usepackage{graphicx}

\usepackage{orcidlink}
\begin{document}

\title{An Efficient Error Estimation Method in Quantum Key Distribution}

\author{\IEEEauthorblockN{1\textsuperscript{st} Yingjian Wang~\orcidlink{0009-0005-3492-7992}}
\IEEEauthorblockA{\textit{Deutsche Telekom Chair of Communication Networks} \\
\textit{Dresden University of Technology}\\
Dresden, Germany \\
yingjian.wang@tu-dresden.de}
\\
\IEEEauthorblockN{3\textsuperscript{rd} Buniechukwu Njoku~\orcidlink{0009-0002-0986-9178}}
\IEEEauthorblockA{\textit{Deutsche Telekom Chair of Communication Networks} \\
\textit{Dresden University of Technology}\\
Dresden, Germany \\
buniechukwu\_chidike.njoku@tu-dresden.de}
\\
\IEEEauthorblockN{5\textsuperscript{th} Riccardo Bassoli~\orcidlink{0000-0002-6132-7985}}
\IEEEauthorblockA{\textit{Deutsche Telekom Chair of Communication Networks} \\
\textit{Dresden University of Technology}\\
Dresden, Germany \\
riccardo.bassoli@tu-dresden.de}
\and
\IEEEauthorblockN{2\textsuperscript{nd} Yilun Hai~\orcidlink{0000-0003-0194-2545}}
\IEEEauthorblockA{\textit{Deutsche Telekom Chair of Communication Networks} \\
\textit{Dresden University of Technology}\\
Dresden, Germany \\
yilun.hai@tu-dresden.de}
\\
\IEEEauthorblockN{4\textsuperscript{th} Koteswararao Kondepu~\orcidlink{0000-0003-0184-1218}}
\IEEEauthorblockA{\textit{Department of Computer Science and Engineering} \\
\textit{Indian Institute of Technology Dharwad}\\
Karnataka, India \\
k.kondepu@iitdh.ac.in}
\\
\IEEEauthorblockN{6\textsuperscript{th} Frank H. P. Fitzek~\orcidlink{0000-0001-8469-9573}}
\IEEEauthorblockA{\textit{Deutsche Telekom Chair of Communication Networks} \\
\textit{Dresden University of Technology}\\
Dresden, Germany \\
frank.fitzek@tu-dresden.de}

\thanks{The authors acknowledge the financial support by the German Research Foundation (DFG, Deutsche Forschungsgemeinschaft) as part of Germany’s Excellence Strategy – EXC 2050/1 – Project ID 390696704 – Cluster of Excellence “Centre for Tactile Internet with Human-in-the-Loop” (CeTI) of Technische Universität Dresden; by the Federal Ministry of Education and Research of Germany in the programme of “Souverän. Digital. Vernetzt.”. Joint project 6G-life, project identification number: 16KISK001K; by the Federal Ministry of Education and Research of Germany in the project 6G-QuaS, project identification number: 16KISQ120; by the Federal Ministry of Education and Research of Germany in the project QD-CamNetz, project identification number: 16KISQ076K.}}

\maketitle

\begin{abstract}
Error estimation is an important step for error correction in quantum key distribution. Traditional error estimation methods require sacrificing a part of the sifted key, forcing a trade-off between the accuracy of error estimation and the size of the partial sifted key to be used and discarded. In this paper, we propose a hybrid approach that aims to preserve the entire sifted key after error estimation while preventing Eve from gaining any advantage. The entire sifted key, modified and extended by our proposed method, is sent for error estimation in a public channel. Although accessible to an eavesdropper, the modified and extended sifted key ensures that the number of attempts to crack it remains the same as when no information is leaked. The entire sifted key is preserved for subsequent procedures, indicating the efficient utilization of quantum resources.
\end{abstract}
\begin{IEEEkeywords}
Quantum key distribution, error estimation, quantum bit error ratio, eavesdropping, bit flipping.
\end{IEEEkeywords}

\section{Introduction}

Quantum key distribution (QKD) was introduced as an alternative key distribution method to achieve information-theoretic security. QKD protocols are typically classified based on the use of discrete variables or continuous variables to encode information, resulting in discrete variable QKD (DV-QKD), such as BB84 \cite{BENNETT20147}, and continuous variable QKD (CV-QKD), such as GG02 \cite{Grosshans_2003}. In general, in a QKD protocol, the sender Alice prepares the raw key and encodes it into quantum information. The raw key is transmitted through a quantum channel and measured by Bob with a sequence of randomly selected measuring bases for DV-QKD or quadrature measurements for CV-QKD. After exchanging the information of the measuring sequence, Alice and Bob are able to agree on a pair of sifted keys. Considering the presence of noise, such as imperfections in the channel, the efficiency of the equipment and the possibility of eavesdropping, the pair of sifted keys cannot be guaranteed to be identical. In practice, Alice and Bob usually exchange a portion of their initial sifted keys to assess parameters such as the quantum bit error ratio (QBER) \cite{Lu_2017,Mehic2020,doi:10.1177/1548512913503418}, determining the presence of eavesdropping, and determining whether the channel noise meets the requirements, in order to decide whether to proceed to the next step, key distillation, or not.

We propose a hybrid method for error estimation, where the sender introduces an additional bit flipping rate to the sifted key and wraps the partially flipped sifted key with random bits. This `seasoned' sifted key is sent through a classical channel to the receiving party. The receiver attempts to estimate the QBER through computation using the received `seasoned' sifted key and the local sifted key. Meanwhile, even if Eve intercepts this `seasoned' sifted key, she cannot reconstruct the original sifted key. The entire original sifted key is thus preserved for the following procedures.

\section{Motivation}

Error estimation is an essential preliminary step for error correction. Parity information related to the sifted key is used for the early CASCADE method \cite{10.1007/3-540-48285-7_35}, with the sacrifice of partial key information and a mess of communication rounds. The current popular error correction methods are usually based on LDPC \cite{9376906,Mao2019,Milicevic2018}, requiring accurate information about the QBER in advance to determine the parameters of the error correction codes, leading to better performance \cite{Maroy2012ErrorEE,Gao2019MultimatrixEE}. While a portion of sifted key is exchanged to estimate the QBER in the traditional method, promising works based on blind information reconciliation, which requires more communication rounds, are proposed \cite{wang2018high,Kiktenko2016SymmetricBI,Liu2020}. E.g., in \cite{wang2018high}, the sifted key is divided into blocks and the first block is used as a pathfinder without knowing the specific quantum channel parameters. Although the reconciled information from the first block can assist in error estimation for subsequent blocks, the absence of QBER in the reconciliation of the first block introduces uncertainty and the potential consumption of the regarding sifted key. Meanwhile, some approaches can only be employed in certain protocols \cite{Lu_2017,wang2018high}. Thus, it's beneficial to design an error estimation method that can:

\begin{itemize}
    \item reduce the consumption of sifted key as much as possible,
    \item accurately estimate the QBER as much as possible,
    \item keep the computational complexity within a reasonable range,
    \item be employed in a generalized scenario.
\end{itemize}

\section{A Hybrid Approach}

In this section, we propose the noisy key method and the bit flipping method as appetizers, and introduce the hybrid method derived from them. They are protocol-agnostic methods based on the traditional method and use the entire sifted key. Some assumptions are made in the paper. The quantum channel is a binary symmetric channel. For convenience, we assume that information for the error estimation, such as the partial sifted key, is sent from Alice to Bob. We also consider the classical channel to be perfect, as its characteristics are not our concern.

\subsection{Noisy key method}
One feasible solution to reduce the leaked information in error estimation is to wrap the sifted key with additional random bits to decrease Eve's chance to tell the sifted bits from the noise bits, while introducing a few uncertainties for error estimation on Bob's side as well. This wrapped information is called a noisy key. 

\begin{figure}[]
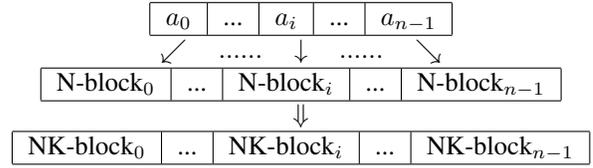

    \centering
    \begin{tabular}{|c|c|c|c|c|}
    \hline
         $a_0$&...&$a_i$&...&$a_{n-1}$  \\
         \hline
    \end{tabular}
    \\
    \begin{tabular}{c c c c c }
         $\swarrow$& $......$ &$\downarrow$& $......$&$\searrow$  \\
    \end{tabular}
    \\
    \begin{tabular}{|c|c|c|c|c|}
    \hline
         $\text{N-block}_0$&...&$\text{N-block}_i$&...&$\text{N-block}_{n-1} $\\
         \hline
    \end{tabular}
    \\
$\Downarrow$
        \\
    \begin{tabular}{|c|c|c|c|c|}
    \hline
         $\text{NK-block}_0$&...&$\text{NK-block}_i$&...&$\text{NK-block}_{n-1} $\\
         \hline
    \end{tabular}

    \caption{The construction of a level-$m$ noisy key.}
    \label{fig-noisy-key}
\end{figure}

We first have insight into the definition of the noise used to construct the noisy key. A string of noise bits of level-$1$ regards a sifted key of length $n$ is a string of random bits of length $n$. Inductively, a string of noise bits of level-$m$ regards a sifted key of length $n$ is a combination of $m$ strings of random bits of length $n$. A noisy key of level-$m$ regards a sifted key of length $n$ is the wrapping of the sifted key and its level-$m$ noise. To be more precise, the $i$-th bit of the sifted key is randomly inserted in one position of the string constructed by the $i$-th bits of all $m$ noise strings of the level-$m$ noise. Here, the position means the position between two adjacent bits of the noise, as well as the position in front of the first bit of the noise and the position after the last bit of the noise. There are $m+1$ possible positions for the $i$-th bit of the sifted key to be inserted. If we denote the string of the sifted key as $(a_0, a_1, ... , a_{n-1})$ and the string of its corresponding level-$m$ noise as $x_0, x_1, ..., x_{n\cdot m -1}$, the $m+1$ possible inserting positions for $a_i$ are the position in front of $x_{i\cdot m}$, the position behind $x_{(i+1)\cdot m-1}$, and the intervals between $x_{i\cdot m}$ and $x_{i\cdot m+1}$, ..., $x_{(i+1)\cdot m-2}$ and $x_{(i+1)\cdot m-1}$. Thus, the sifted key is wrapped with its relevant level-$m$ noise, resulting in a noisy key with the length of $n\cdot (m+1)$. For simplicity, we call the bits $(x_{i\cdot m},..., x_{(i+1)\cdot m-1})$, viz., the $i$-th bits of $m$ noise strings, the $i$-th $\text{N-block}$ of the {level-$m$} noise. $\text{N-block}_i$ after inserting $a_i$ is called the $i$-th $\text{NK-block}$ of the noisy key. Fig. \ref{fig-noisy-key} demonstrates the construction of a level-$m$ noisy key of a length $n$ sifted key and its level-$m$ noise. The down arrows represent the insertions and the double down arrow indicates the relations between the insertions and the result.

Alice's use of noisy keys instead of sifted keys during transmission provides a certain degree of defense against Eve's eavesdropping. At the same time, Bob can still perform error estimation without the need to identify the sifted key.

For Bob, when he receives a level-$m$ noisy key, he uses his own sifted key, of length $n$, to perform error estimation. Specifically, each bit of Bob's sifted key is compared with the corresponding bits in the noisy key, viz., every bit in its corresponding $\text{NK-block}$. If they differ, a counter is incremented, resulting in a total error count, denoted as $\sharp$error. It is worth noting that this procedure can be implemented in a parallel way, with which the computation duration is decreased. Bob's sifted key is compared with the noisy key a total of $n\cdot (m+1)$ times, yielding the bit error ratio $BER_{\text{noisy\_ key}}$ of the noisy key with respect to Bob's sifted key, calculated as $\frac{\sharp \text{error}}{n(m+1)}$.

Since the noisy key is a combination of Alice's sifted key and $m\cdot n$ random noise bits, $BER_{\text{noisy\_ key}}$ can be decomposed into $\frac{QBER_\text{sifted\_key} + m\cdot BER_\text{noise}}{m+1}$. Here, $QBER_{\text{sifted\_key}}$ stands for the part that is computed using the sifted keys of both Alice and Bob and it contributes to restoring the original QBER. $BER_{\text{noise}}$ stands for the part that is computed using Bob's sifted key and the level-$m$ noise. Since the noise has $m$ times the length of Bob's sifted key and Bob uses each bit to check the difference for its corresponding $\text{NK-block}$, $BER_\text{noise}$ has $m$ as its weight. Both the noise and Bob's sifted key are random bits, two strings of random bits are expected to have $50\%$ of the bits to be different, this probability obeys binomial distribution. Thus, estimating $BER_\text{noise}$ as $50\%$, calculating $BER_{\text{noisy\_ key}}$ can be specifically abbreviated as $\frac{QBER_\text{sifted\_key} + m/2}{m+1}$.

It should be noted that $QBER_\text{sifted\_key}$ is not explicitly obtained. Nevertheless, since $BER_{\text{noisy\_key}}$ is directly obtained by calculation, we can derive that 
\begin{equation}
    QBER_\text{sifted\_key}=(m+1)\cdot BER_{\text{noisy\_ key}} -  m\cdot BER_{\text{noise}}
    \label{eqqber}
\end{equation}

 and set $BER_{\text{noise}}$ to $50\%$ to for an approximate estimation. The inserting location of each bit from Alice's sifted key in the noisy key does not affect the calculation of QBER. On the contrary, for Eve, the specific location of each bit from Alice's sifted key in the noisy key is crucial.
If Eve cannot differentiate between sifted bits and noise bits, she must attempt both cases, $0$ and $1$, rendering it impossible for her to gain any advantage.

Since we focus more on the case that the QBER is in an acceptable range, for example, no more than $40\%$ \cite{9376906}, the accuracy when QBER is low is more important than the accuracy when QBER is high. Too much QBER means an unreliable channel and extra effort for decoding. We suggest calculating the BER of Alice's sifted key and its level-$m$ noise as the value for $BER_{\text{noise}}$, rather than $50\%$, to estimate the QBER on Bob's side. This kind of approximation is supposed to result in a more accurate estimation when QBER is low.

There is an expected $50\%$ overlap of bits between two equilong random strings. This implies that for a level-m noisy key, an expected $\frac{100}{2^m}\%$ of the bits are identical, meaning that these $\text{NK-block}$s have only one kind of bits, $1$ or $0$. In this scenario, Eve unconditionally knows the values of the sifted bits in the above mentioned $\text{NK-block}$s and benefits from it. 
\begin{equation}
    p_{\text{leaked}}=\frac{100}{2^m}\%
\end{equation}
    
is thus called the unconditional leaking rate of the level-m noisy key.

Another point to note is that Eve has an effective cracking strategy for level-$m$ noisy keys when $m$ is even. She always chooses the value that appears the most frequently in each $\text{NK-block}$. This strategy guarantees Eve a 75\% probability to guess correctly when $m=2$ as shown in TABLE \ref{eve}. The chance is $\frac{2^{m}+C_m^{\frac{m}{2}}}{2^{m+1}}$ when talking about a general even number $m$ and the chance approaches 50\% as $m$ grows.

\begin{table}[]
    \centering
    \caption{An example of Eve's guess strategy when sifted bit is 0 and noise is level-2.}
    \begin{tabular}{c|c|c|c|c}
    Sifted bit&Noise bits&Probability&Eve's guess&Hit\\
    \hline
         0&00&$25\%$&0&$\checkmark$  \\
         0&01&$25\%$&0&$\checkmark$  \\
         0&10&$25\%$&0&$\checkmark$  \\
         0&11&$25\%$&1&$\usym{2613}$  \\
    \end{tabular}
    
    \label{eve}
\end{table}

\subsection{Bit flipping method}
A feasible solution to address the information leakage caused by the noisy key is to randomly select and invert some of the sifted bits to be transmitted. Intuitively, this will introduce further uncertainty to error estimation, while leading to an increase in Eve's search efforts. With the introduction of the flipping of $k$ bits, the formula for calculating the QBER for sifted key has also been changed. We denote the original bit error probability by $p_\text{error}$, the bit error probability after bit flipping as $p_\text{f\_error}$, and bit flipping rate as $p_\text{flip}$. We estimate
\begin{equation}
p_\text{f\_error}=p_\text{flip}\cdot (1-p_\text{error})+(1-p_\text{flip})\cdot p_\text{error}.
\end{equation}

When we calculate and get $QBER_\text{flip}$, viz., the specific value for $p_\text{f\_error}$, we can attempt to recover the original $QBER$, viz., the specific value for $p_\text{error}$, by using 
\begin{equation}
    QBER = \frac{QBER_\text{flip}-p_\text{flip}}{1-2p_\text{flip}}.
    \label{eqflip}
\end{equation}

While $QBER_\text{flip}$ preserves relevance to the original $QBER$, extra effort to crack the origin sifted key is introduced with a partially flipped sifted key. Flipping $k$ bits out of $n$ would lead to $C^k_n$ combinations. While $C^k_n$ can make a limited contribution to reach the maximum attempts to crack $n$ bits, $2^n$, it is beneficial to combine the bit flipping method with the noisy key method.

\subsection{Hybrid method}
Although the noisy key method can achieve the original cracking complexity with a large $m$, introducing too many noise bits can result in overhead traffic in the classical channel. Introducing the bit flipping method to the noisy packet can improve its security. Theoretically, for a sifted key of length $n$, let \(p_{\text{leaked}}\) be the unconditional hit rate for Eve due to the interception of the corresponding noisy key produced by the noisy key method. In this case, Eve needs \(2^{n\cdot(1-p_{\text{leaked}})}\) attempts to obtain all the sifted bits. If we use the bit flipping method before using the noisy key method and set the bit flipping ratio to \(p_{\text{leaked}}\), Eve's search attempts can be described as 

\begin{align}
Attempts_{\text{Eve}} & = 2^{n(1 - p_{\text{leaked}})} \cdot \left(C^{0}_{n \cdot p_{\text{leaked}}} + C^{1}_{n \cdot p_{\text{leaked}}} + \ldots \right. \notag \\
& \left. + C^{n \cdot p_{\text{leaked}}}_{n \cdot p_{\text{leaked}}}\right).
\label{eqh}
\end{align}

The second term indicates the case where Eve has to determine how many bits from the leaked bits have already been flipped. Since we set the flipping ratio to \( p_{\text{leaked}} \), Eve must check all possible scenarios, from no leaked bit being flipped to all leaked bits being flipped. This results in a total search of \( 2^{n \cdot p_{\text{leaked}}} \) possibilities. Thus, we expect \( Attempts_{\text{Eve}} \) to be \( 2^n \), meaning that Eve is not able to gain an advantage with the knowledge of the so-called leaked bits.

In this case, it can be theoretically ensured that there is the same size of cracking search space even when the noisy key is leaked. Considering the expected amount of flipped bits appearing in the identical NK-blocks follows a binomial distribution, we should let $p_{\text{flipped}} > p_\text{leaked}$.
\begin{figure*}
    \centering

        \subfloat[Sifted key length is 100 bits.]{
        \includegraphics[scale=0.5]{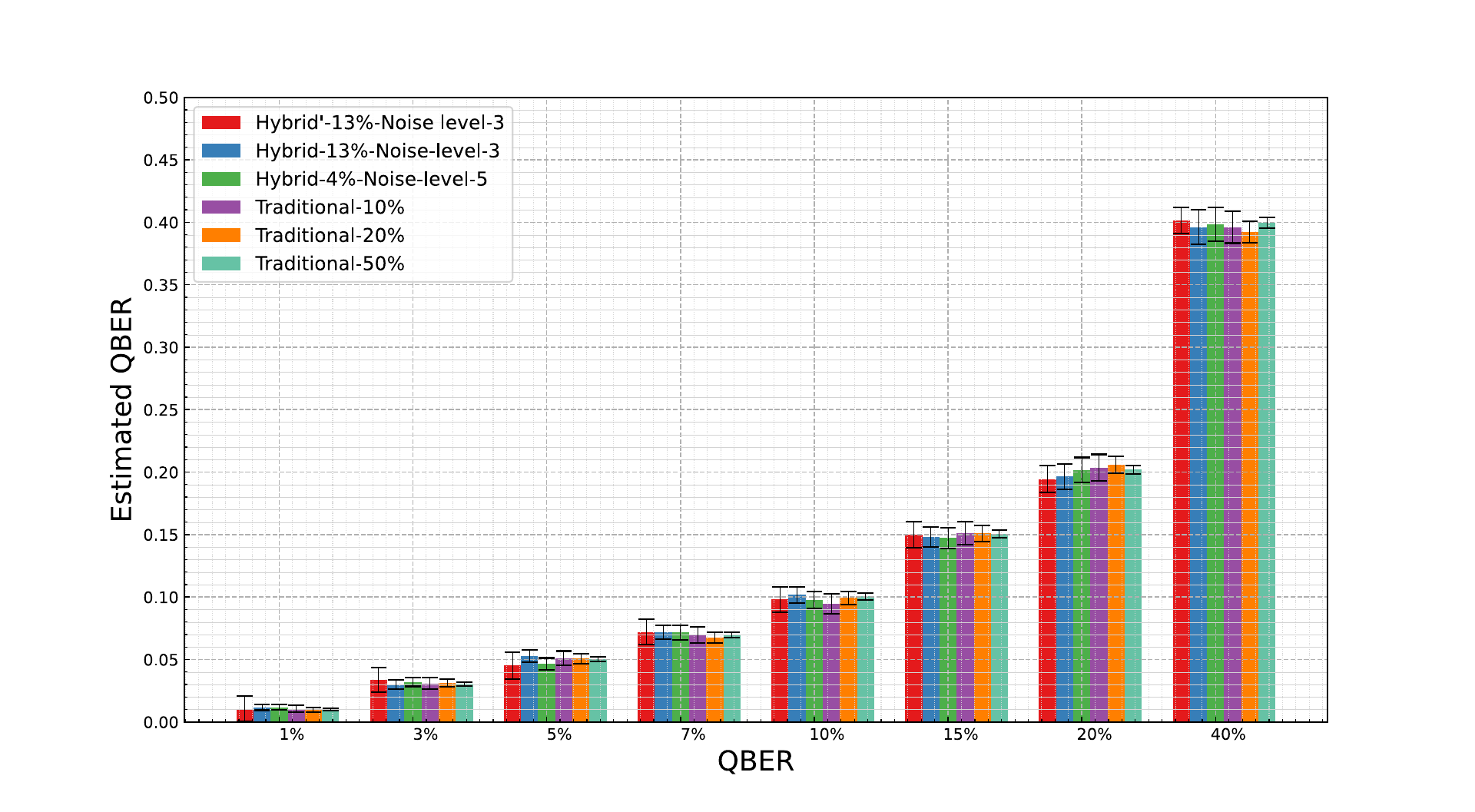}
        \label{fig:sub1}
    }
    \\
    \subfloat[Sifted key length is 500 bits.]{
        \includegraphics[scale=0.5]{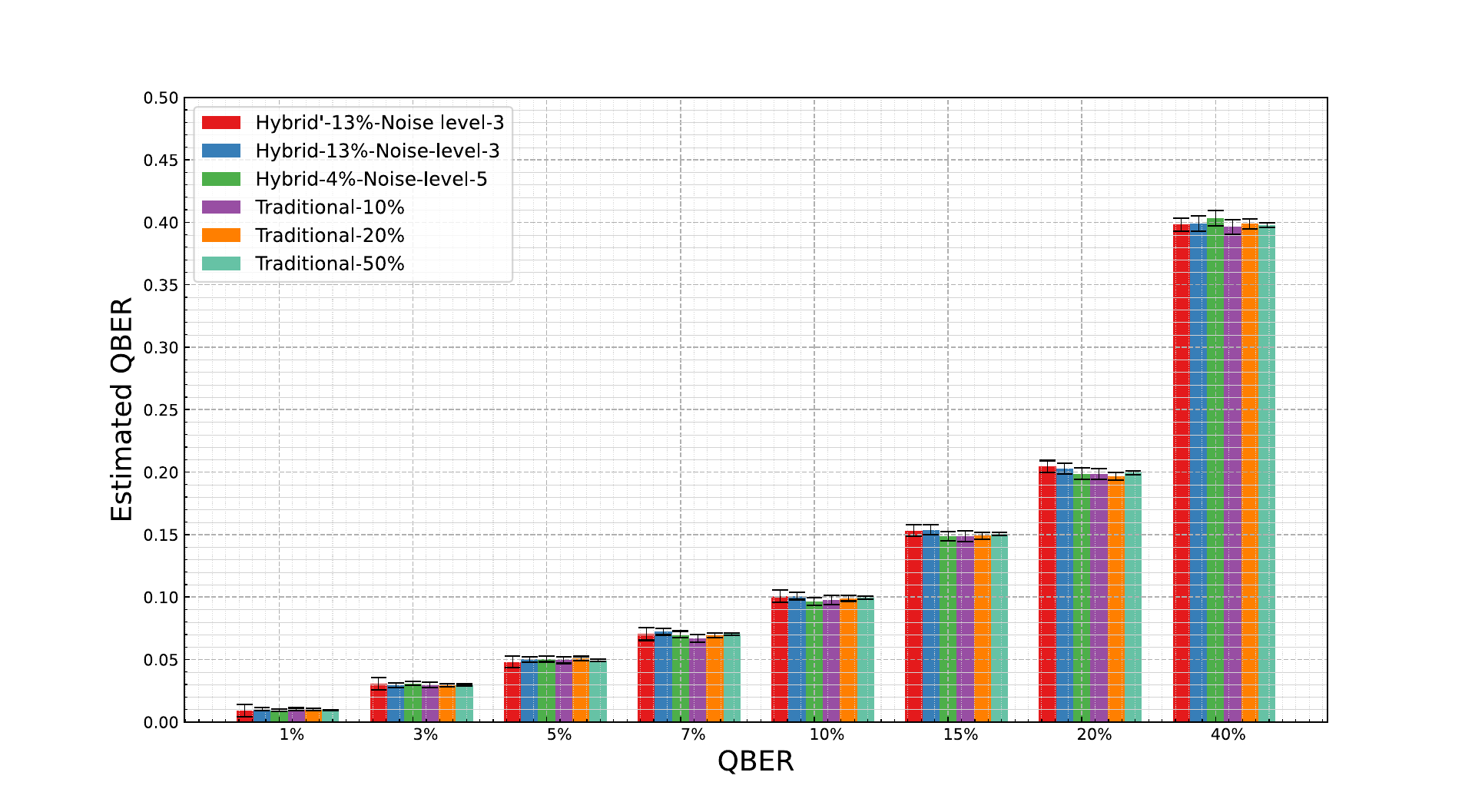}
        \label{fig:sub2}
    }

    \caption{Performances for different methods with different parameters under different QBERs and key lengths.}
    \label{fig:3d}
\end{figure*}

The proposed approach, referred to as the hybrid method, does not increase time complexity compared to the classical method when calculating the QBER. The procedure for calculating QBER in the hybrid method is straightforward: first, apply formula \ref{eqqber}, and then use the result from formula \ref{eqqber} to apply formula \ref{eqflip}. Nevertheless, in practical applications, since the hybrid method introduces more operations, the operational time of the hybrid method is higher compared to the traditional method. It is worth noting that using the noisy flipped key, abbreviated as the noisy key in the context of hybrid approach, has been proven to be secure. This implies that both Alice and Bob can act as producers and senders of such key. In other words, researchers can decide either Alice or Bob would be the producer and sender of the noisy key to achieve an optimal communication sequence in their use cases. One should also notice that the amount of data transmitted in regarding communication using our proposed method differs from the traditional approach. The size of the noisy key is $(m+1)n$ bits, where $n$ represents the length of the sifted key, and $m$ denotes the noise level. This results in additional traffic, considering that the traffic in the traditional method is usually no more than $n/2$ bits.

\section{Results and Discussion}

In this section, we will simulate and analyze the traditional method and the hybrid method under different QBERs. The simulation results are presented in Fig. \ref{fig:3d}, considering that the sifted key lengths are set to 100 and 500 bits and shown in sub-figures \ref{fig:sub1} and \ref{fig:sub2}, respectively. For each case, we consider 500 samples and take their mean as the value, shown as the height of the bars with different colors. 
In each sub-figure, e.g., in \ref{fig:sub1} the horizontal axis represents the real QBER of the quantum channel from 1\% over 3\%, 5\%, 7\% and 10\%, representing the performances of common reliable quantum channels, to 15\%, 20\% and 40\%, demonstrating the performances of quantum channels under comparable risky conditions, while the vertical axis represents the estimated results by using different methods and parameters shown in the legend with different colors. Specifically, `Hybrid' or `Traditional' demonstrate the method used. The percentage stands for the bit flipping ratio for the hybrid method or the percentage of bits consumed for the traditional method, respectively. `Noise-level-3' and `Noise-level-5' stands for the noise level specified in the previous section. We should notice that the `Hybrid' method uses the $BER_\text{noise}$ from Alice, while the `Hybrid$'$' method uses `$50\%$', to calculate the QBER depicted in formula \ref{eqqber}. The error lines at the upper part of each bar represent the 95\% confidence intervals.

The selected SNR range is based on both common low BER scenarios and extreme cases with high BER. This choice is made to facilitate a more comprehensive comparison between our proposed method and traditional methods across different application contexts. We consider the SNR to be no more than 40\%, since for a quantum channel whose QBER is greater than 40\%, for example 50\%, meaning that half of the messages passing through this channel are expected to be incorrect, would be considered unsafe and the current communication should be discarded or some other operations should be performed to improve the communication quality.

Insight into Fig. \ref{fig:3d}, the traditional method consistently manifests performance stability across all two sub-figures, notably when employing 50\% of the key for error estimation. As more practical cases, compared to `Traditional-50\%', `Traditional-20\%' and `Traditional-10\%' maintain relatively consistent means while also increasing the corresponding confidence intervals. Our hybrid approaches exhibit efficacy comparable to `Traditional-20\%' and `Traditional-10\%' in QBER estimation, maintaining accuracy within an acceptable range as delineated by the confidence interval. To be more precise, in the context of the shorter key length, 100 bits in \ref{fig:sub1}, `hybrid-13\%-Noise-Level-3' and `hybrid-4\%-Noise-Level-5' underscore the potential to supplant `Traditional-10\%' in QBER estimation concerning mean value and stability with acceptable error margins when QBER is lower than 20\%, while `hybrid$'$-13\%-Noise-Level-3' demonstrates superiority at QBER of 40\%. However, it is not a wise choice to use `hybrid$'$-13\%-Noise-Level-3' when SNR is lower than 15\%, the assumption of using 50\% to calculate  QBER shows its weakness under low QBER condition as demonstrated in the previous section. In terms of a longer sifted key length, all methods maintained their performance observed with a sifted key length of 100 bits, demonstrating the stability of each method's inherent behavior. Additionally, with the increase in sifted key length, the impact of random phenomena, such as noise, tends to average out. This is reflected in the more consistent occurrence of the estimated QBER and the narrowing of confidence intervals. As a conclusion, it is beneficial to use the 'Hybrid' method and the 'Hybrid$'$'method according to different BER scenarios. 

On the other hand, our proposed method significantly reduces the consumption of the sifted key in error estimation compared to traditional approaches. As shown in Table \ref{tab2}, our method preserves the entire sifted key for subsequent steps, while the traditional method retains only a portion of the sifted key. To be more precise, taking 'Traditional-10\%' as a competitor, since it has comparable performance with our proposed approach, 10\% of the sifted key is exposed and consumed for BER estimation. This means that, in the case where privacy amplification is not applied and the sifted key itself is the final key, 'Traditional-10\%' generates 10\% fewer final key bits per QKD procedure compared to 'Hybrid/Hybrid$'$'. For the case that privacy amplification is applied, since the sifted keys generated by 'traditional-10\%' methods already meet the collision resistance requirements for privacy amplification, the extra portion of the sifted key produced by the 'Hybrid/Hybrid$'$' method should be retained in the buffer. Theoretically, after nine rounds of QKD, the extra portions retained in the buffer from these nine rounds could be combined to form a new sifted key. Regarding to the complexity, as discussed in the previous section, despite introducing additional computations, our proposed method introduces no new function for data processing and maintains the same time complexity as traditional methods, meaning that the computation is not a bottleneck in the QKD process. Meanwhile, as discussed in the previous section, a hybrid method, whose $p_\text{flipped}$ is greater than $p_\text{leaked}$, is robust to the leak of its noisy flipped sifted key. This ensures that Eve's attempts to crack the original sifted key still amount to $2^{|\text{sifted\_key}|}$.
\begin{table}[]
    \centering
    \caption{The comparison among different methods.}
    \begin{tabular}{|c|c|c|}
    \hline
    {Methods}&{Hybrid$'$/Hybrid}&{Traditional-10\%$\sim$50\%}\\
    \hline
         {Sifted key preserved}&$100\%$&$50\%\sim 90\%$  \\
         \hline
         {Time complexity}&$O(n)$&$O(n)$  \\
    \hline
             {Attempts to crack}&$2^{|\text{sifted\_key}|}$&$2^{|\text{sifted\_key}|}$ \\
    \hline
    \end{tabular}
    
    \label{tab2}
\end{table}
\section{Conclusion}

We proposed a hybrid approach for error estimation in a QKD procedure. All sifted keys are used without information leakage, thanks to the noisy key and bit flipping method. The proposed approach outperforms the traditional method using 10\% of the sifted key while no sifted key is consumed, achieves more efficient utilization of quantum resources. Our proposed method performs well with sifted key lengths as low as 100 bits, signifying its applicability to a wider range of scenarios. For instance, in the block-based error estimation mentioned in \cite{Kiktenko2016SymmetricBI}, where block lengths can reach over thousands of bits, our method can be integrated without causing a loss in the sifted key. Our proposed method has the same time complexity as the traditional method and allows for concurrent calculation to estimate the QBER, which further decreases the computing duration in time-sensitive scenarios.

\bibliographystyle{IEEEtran}
\bibliography{IEEEabrv,main}

\begin{thebibliography}{10}
\providecommand{\url}[1]{#1}
\csname url@samestyle\endcsname
\providecommand{\newblock}{\relax}
\providecommand{\bibinfo}[2]{#2}
\providecommand{\BIBentrySTDinterwordspacing}{\spaceskip=0pt\relax}
\providecommand{\BIBentryALTinterwordstretchfactor}{4}
\providecommand{\BIBentryALTinterwordspacing}{\spaceskip=\fontdimen2\font plus
\BIBentryALTinterwordstretchfactor\fontdimen3\font minus \fontdimen4\font\relax}
\providecommand{\BIBforeignlanguage}[2]{{%
\expandafter\ifx\csname l@#1\endcsname\relax
\typeout{** WARNING: IEEEtran.bst: No hyphenation pattern has been}%
\typeout{** loaded for the language `#1'. Using the pattern for}%
\typeout{** the default language instead.}%
\else
\language=\csname l@#1\endcsname
\fi
#2}}
\providecommand{\BIBdecl}{\relax}
\BIBdecl

\bibitem{BENNETT20147}
\BIBentryALTinterwordspacing
C.~H. Bennett and G.~Brassard, ``Quantum cryptography: Public key distribution and coin tossing,'' \emph{Theoretical Computer Science}, vol. 560, pp. 7--11, 2014, theoretical Aspects of Quantum Cryptography -- celebrating 30 years of BB84. [Online]. Available: \url{https://www.sciencedirect.com/science/article/pii/S0304397514004241}
\BIBentrySTDinterwordspacing

\bibitem{Grosshans_2003}
\BIBentryALTinterwordspacing
F.~Grosshans, G.~V. Assche, J.~Wenger, R.~Brouri, N.~J. Cerf, and P.~Grangier, ``Quantum key distribution using gaussian-modulated coherent states,'' \emph{Nature}, vol. 421, no. 6920, pp. 238--241, jan 2003. [Online]. Available: \url{https://doi.org/10.1038%2Fnature01289}
\BIBentrySTDinterwordspacing

\bibitem{Lu_2017}
\BIBentryALTinterwordspacing
Z.~Lu, J.-H. Shi, and F.-G. Li, ``Error rate estimation in quantum key distribution with finite resources,'' \emph{Communications in Theoretical Physics}, vol.~67, no.~4, p. 360, apr 2017. [Online]. Available: \url{https://dx.doi.org/10.1088/0253-6102/67/4/360}
\BIBentrySTDinterwordspacing

\bibitem{Mehic2020}
\BIBentryALTinterwordspacing
M.~Mehic, M.~Niemiec, H.~Siljak, and M.~Voznak, \emph{Error Reconciliation in Quantum Key Distribution Protocols}.\hskip 1em plus 0.5em minus 0.4em\relax Cham: Springer International Publishing, 2020, pp. 222--236. [Online]. Available: \url{https://doi.org/10.1007/978-3-030-47361-7_11}
\BIBentrySTDinterwordspacing

\bibitem{doi:10.1177/1548512913503418}
\BIBentryALTinterwordspacing
J.~S. Johnson, M.~R. Grimaila, J.~W. Humphries, and G.~B. Baumgartner, ``An analysis of error reconciliation protocols used in quantum key distribution systems,'' \emph{The Journal of Defense Modeling and Simulation}, vol.~12, no.~3, pp. 217--227, 2015. [Online]. Available: \url{https://doi.org/10.1177/1548512913503418}
\BIBentrySTDinterwordspacing

\bibitem{10.1007/3-540-48285-7_35}
G.~Brassard and L.~Salvail, ``Secret-key reconciliation by public discussion,'' in \emph{Advances in Cryptology --- EUROCRYPT '93}, T.~Helleseth, Ed.\hskip 1em plus 0.5em minus 0.4em\relax Berlin, Heidelberg: Springer Berlin Heidelberg, 1994, pp. 410--423.

\bibitem{9376906}
\BIBentryALTinterwordspacing
S.-S. Yang, J.-Q. Liu, Z.-G. Lu, Z.-L. Bai, X.-Y. Wang, and Y.-M. Li, ``An fpga-based ldpc decoder with ultra-long codes for continuous-variable quantum key distribution,'' \emph{IEEE Access}, vol.~9, pp. 47\,687--47\,697, 2021. [Online]. Available: \url{https://doi.org/10.1109/ACCESS.2021.3065776}
\BIBentrySTDinterwordspacing

\bibitem{Mao2019}
\BIBentryALTinterwordspacing
H.~Mao, Q.~Li, Q.~Han, and H.~Guo, ``High-throughput and low-cost ldpc reconciliation for quantum key distribution,'' \emph{Quantum Information Processing}, vol.~18, no.~7, p. 232, Jun 2019. [Online]. Available: \url{https://doi.org/10.1007/s11128-019-2342-2}
\BIBentrySTDinterwordspacing

\bibitem{Milicevic2018}
\BIBentryALTinterwordspacing
M.~Milicevic, C.~Feng, L.~M. Zhang, and P.~G. Gulak, ``Quasi-cyclic multi-edge ldpc codes for long-distance quantum cryptography,'' \emph{npj Quantum Information}, vol.~4, no.~1, p.~21, Apr 2018. [Online]. Available: \url{https://doi.org/10.1038/s41534-018-0070-6}
\BIBentrySTDinterwordspacing

\bibitem{Maroy2012ErrorEE}
\BIBentryALTinterwordspacing
O.~Maroy, M.~Gudmundsen, L.~Lydersen, and J.~Skaar, ``Error estimation, error correction and verification in quantum key distribution,'' \emph{IET Inf. Secur.}, vol.~8, pp. 277--282, 2012. [Online]. Available: \url{https://api.semanticscholar.org/CorpusID:3191656}
\BIBentrySTDinterwordspacing

\bibitem{Gao2019MultimatrixEE}
\BIBentryALTinterwordspacing
C.~Gao, D.~Jiang, Y.~Guo, and L.~Chen, ``Multi-matrix error estimation and reconciliation for quantum key distribution.'' \emph{Optics express}, vol. 27 10, pp. 14\,545--14\,566, 2019. [Online]. Available: \url{https://api.semanticscholar.org/CorpusID:155169643}
\BIBentrySTDinterwordspacing

\bibitem{wang2018high}
\BIBentryALTinterwordspacing
X.~Wang, Y.~Zhang, S.~Yu, and H.~Guo, ``High efficiency postprocessing for continuous-variable quantum key distribution: using all raw keys for parameter estimation and key extraction,'' \emph{Quantum Information Processing}, vol.~18, no.~9, p. 264, Jul 2019. [Online]. Available: \url{https://doi.org/10.1007/s11128-019-2381-8}
\BIBentrySTDinterwordspacing

\bibitem{Kiktenko2016SymmetricBI}
\BIBentryALTinterwordspacing
E.~O. Kiktenko, A.~S. Trushechkin, C.~C.~W. Lim, Y.~V. Kurochkin, and A.~K. Fedorov, ``Symmetric blind information reconciliation for quantum key distribution,'' \emph{Phys. Rev. Appl.}, vol.~8, p. 044017, Oct 2017. [Online]. Available: \url{https://link.aps.org/doi/10.1103/PhysRevApplied.8.044017}
\BIBentrySTDinterwordspacing

\bibitem{Liu2020}
\BIBentryALTinterwordspacing
Z.~Liu, Z.~Wu, and A.~Huang, ``Blind information reconciliation with variable step sizes for quantum key distribution,'' \emph{Scientific Reports}, vol.~10, no.~1, p. 171, Jan 2020. [Online]. Available: \url{https://doi.org/10.1038/s41598-019-56637-y}
\BIBentrySTDinterwordspacing

\end{thebibliography}

\end{document}